\documentclass[9pt,twocolumn,twoside]{pnas-new}

\templatetype{pnasresearcharticle} 

\title{Artifact magnification on deepfake videos increases human detection and subjective confidence}

\author[a]{Emilie L. Josephs}
\author[a]{Camilo L. Fosco} 
\author[a]{Aude Oliva}

\affil[a]{Computer Science and Artificial Intelligence Lab, MIT, 77 Massachusetts Ave, Cambridge, MA 02139, USA}

\leadauthor{Josephs} 

\significancestatement{It is important to understand where and when someone might fall for a deepfake. Previous experiments which measure this typically use settings that don’t reflect real-world browsing experiences. Here, we identify four conditions that can exist during typical browsing (low prevalence, short or low quality videos, and divided attention), and found that they all increased users’ likelihood of falling for a deepfake. This suggests that users are more vulnerable to deepfakes than previously thought. We also explored ways to flag fake videos for users. We found that using text to tell people that a video is fake increases performance, but showing them that the video was fake by amplifying the artifacts was even more convincing. }

\authorcontributions{Author contributions: E.J., C.F. and A.O designed research, E.J performed research, analyzed data and wrote paper, C.F. contributed computer vision tools, A.O. and C.F. provided insight throughout and editing during the writing process}
\authordeclaration{The authors have no competing interests to declare.}
\correspondingauthor{\textsuperscript{2}To whom correspondence should be addressed. E-mail: ejosephs@mit.edu}


\begin{abstract}
The development of technologies for easily and automatically falsifying video has raised practical questions about people’s ability to detect false information online. How vulnerable are people to deepfake videos? What technologies can be applied to boost their performance? Human susceptibility to deepfake videos is typically measured in laboratory settings, which do not reflect the challenges of real-world browsing. In typical browsing, deepfakes are rare, engagement with the video may be short, participants may be distracted, or the video streaming quality may be degraded. Here, we tested deepfake detection under these ecological viewing conditions, and found that detection was lowered in all cases. Principles from signal detection theory indicated that different viewing conditions affected different dimensions of detection performance. Overall, this suggests that the current literature underestimates people’s susceptibility to deepfakes. Next, we examined how computer vision models might be integrated into users’ decision process to increase accuracy and confidence during deepfake detection. We evaluated the effectiveness of communicating the model’s prediction to the user by amplifying artifacts in fake videos. We found that artifact amplification was highly effective at making fake video distinguishable from real, in a manner that was robust across viewing conditions. Additionally, compared to a traditional text-based prompt, artifact amplification was more convincing: people accepted the model’s suggestion more often, and reported higher final confidence in their model-supported decision, particularly for more challenging videos. Overall, this suggests that visual indicators that cause distortions on fake videos may be highly effective at mitigating the impact of falsified video.
\end{abstract}

\begin{document}

\maketitle
\ifthenelse{\boolean{shortarticle}}{\ifthenelse{\boolean{singlecolumn}}{\abscontentformatted}{\abscontent}}{}

\dropcap{D}eepfakes are increasingly common, increasingly easy to create, and increasingly convincing. “Deepfake” is the colloquial term for images, video, or audio that has been manipulated using deep learning techniques \cite{deepfake_def_MW}. Most typically, these involve manipulating the identity or appearance of a person. They are often harmless, and can have positive applications, helping to create movies, video games, whimsical social media filters, or novel and striking art. However they can be used for malicious purposes, ranging from fraud, impersonation, blackmail, fake news and political propaganda. They present a risk on any platform that supports images or video, including social media, ads, video conferencing platforms, tip lines, or official websites. Their propagation in the modern  information landscape raises at least two major practical and societal questions. How effective are deepfake videos at fooling human observers? What is the best way to warn viewers about deepfakes, and insulate them from the false information they contain? 

Human users are highly susceptible to deepfakes. Deepfake images of faces have reached a point where they are indistinguishable from real images, and may even elicit higher levels of trust and social compliance \cite{nightingale2022ai, tucciarelli2022realness, shen2021study, lago2021more}. Deepfake videos are somewhat less convincing: users can still detect them at above chance levels \cite{groh2022deepfake, kobis2021fooled, korshunov2021subjective, rossler2019faceforensics++, lovato2022diverse}, but detection rates are still well below ceiling. It is impossible to provide a single estimate of video deepfake detection rates, because studies differ in their design and stimuli, but a survey of recent deepfake video detection studies suggest average detection rates in the 60-70\% range \cite{groh2022deepfake, rossler2019faceforensics++, boyd2022value, prasad2022human}. Thus, in a set of 10 fake videos, up to 4 of them might get past a typical user.

Critically, people’s ability to detect deepakes is likely even lower than the current literature suggests. Most recent experiments use settings that can inflate detection rates: participants are informed about deepfakes, responses are untimed, deepakes are abundant, video streaming quality is controlled. During a real-word browsing session, people might not be searching under such ideal conditions. In practice, deepfakes are rare, people are distracted, video quality is variable, and videos can be short. All of these conditions can impair the detection of even the most obvious signals \cite{rich2008we, prasad2022human}. To date, there is little understanding of human deepfake detection varies under more ecologically-valid search conditions. Here, we perform the first systematic analysis of how a range of detection conditions, based on real-world challenges, affect deepfake detection in humans. To anticipate, we find that all conditions we tested reduced the detectability of deepfake videos.

Thus, people’s ability to detect and reject false videos remains low, and some intervention is required to boost it. One suggestion has been to change the motivational state of the viewer, using high-level interventions. However, these methods have not shown significant success. Attempting to motivate participants by teaching them about the harms of deepfakes \cite{kobis2021fooled}, adding financial incentives \cite{kobis2021fooled}, or even eliciting emotional states \cite{groh2022deepfake} have all failed to affect detection rates. A second, emerging direction is to supplement human users with additional information about the video from an independent source. Specifically, recent work has suggested human-AI teaming, where users are given access to a computer vision model that specializes in deepfake detection \cite{boyd2022value, groh2022deepfake, sohrawardi2020defaking}.

Pairing humans with models is an emerging possibility because of rapid advances in deepfake detection by machine learning models. Most models use computer vision methods, detecting signs of tampering in the video such as blending or blurring artifacts \cite{durall2019unmasking, li2019face, li2019zooming, yang2019exposing, li2020identification}, anomalies in the biological signals of the video’s subject (e.g. heartbeat, blinking, teeth) \cite{li2018ictu, ciftci2019fakecatcher, matern2019exploiting}, or inconsistencies in individual-specific features, such as facial, gestural or vocal features  \cite{bohavcek2022protecting, agarwal2020detecting, haliassos2021lips, cozzolino2021id, yang2019exposing, matern2019exploiting}. Another kind of approach relies on authenticating a video based on its metadata, with proposals such as digital watermarking, blockchain-based tracking, and dataset fingerprinting \cite{qureshi2021detecting, neekhara2022facesigns, chan2020combating, alattar2020system, yu2021artificial}. In general, methods for automatically detecting deepfakes are under active development. 

So far, however, teaming humans with AI assistants has met with mixed success. Many of these models are as good as (if not better than) humans \cite{groh2022deepfake, korshunov2021subjective, rossler2019faceforensics++}, and make complementary mistakes \cite{groh2022deepfake, korshunov2021subjective, prasad2022human}. Under the right circumstance, then, humans paired with models can reach higher accuracy than humans alone \cite{boyd2022value}. However, these studies also show surprisingly low model acceptance rates: human users are relatively reluctant to accept model suggestions. In one study, participants who had access to model suggestions only updated their responses 24\% of the time, and only changed their mind 12\% of the time \cite{groh2022deepfake}. Even when models are highly accurate, people are embracing them only some of the time: Boyd et al \cite{boyd2022value} found that teaming humans with a model that is 90\% accurate only yielded final human accuracy of 63\%. Overall, current approaches to human-AI teaming for deepfake detection have significant unrealized potential.

How can we increase viewer's engagement with model suggestions? In traditional approaches, model's predictions are communicated to the user using text. One direction for improvement may be to develop visual indicators that are more intuitive and compelling. Recent efforts tried showing users saliency maps of suspicious video regions, but these did not improve engagement relative to text-based indicators \cite{boyd2022value, malolan2020explainable}. Here, we propose a novel approach to visual indicator design, which relies on motion magnification to amplify artifacts in fake video. 

Artifact amplification is well-suited to deepfake signaling for many reasons. First, it is a highly intuitive signal. It targets and amplifies the same information that humans instinctively use to make an unassisted judgment, like the naturalness of motion and the coherence of the faces \cite{mittal2022gotcha}. It also targets a visual domain that humans are particularly attuned to. Humans are exceptionally sensitive to the proportions of faces \cite{benson1991perception, mauro1992caricature, farah1998special, sinha2006face} and are highly sensitive to unnatural faces (e.g. the uncanny valley effect). Second, it is practical: videos in the current online landscape are already loaded with text and icons (e.g. video playback controls, social media platform watermarks, news tickers and crawlers, closed captions), so additional text may not be very salient compared to distortions over the face itself. Here, we test the validity of artifact amplification for signaling deepfake videos to human observers. 

 Altogether, the present paper makes two broad contributions to the science of human deepfake detection: it advances our understanding of the risks they pose to humans, by charting the detectability of deepfakes across more ecological conditions than previously considered, and explores a novel alternative for mitigating this risk. Taken together, these experiments provide insights to improve the effectiveness of human-AI teaming in deepfake detection.

\section*{Results}

\subsection*{Typical browsing conditions reduce deepfake detection performance}

\begin{figure}
\centering
\includegraphics[width=0.9\linewidth]{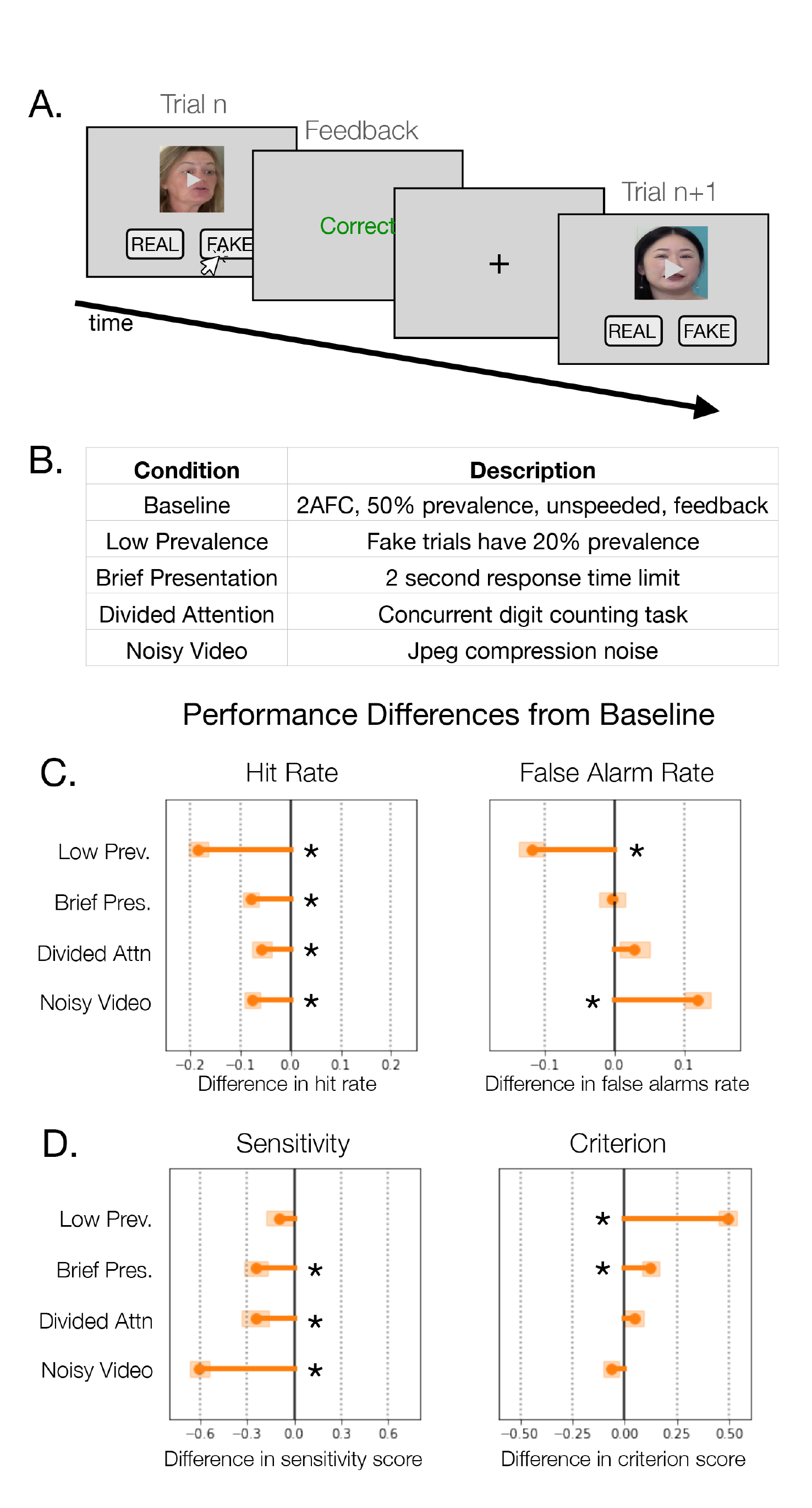}
\caption{Methods, conditions and results of deepfake detection under ecological conditions. A) Baseline procedure: participants viewed one video at a time, and indicated whether they thought it was real or fake. Feedback was given on each trial. B) Summary of the conditions explored. The baseline procedure was modified to accommodate each condition, see Methods. C) Results: changes in hit and false alarm rates relative to baseline. Light orange box indicates standard error of the mean, and stars indicate significance. Baseline hit and false alarm rates were 0.73 and 0.28 respectively. D) Results: change in sensitivity and criterion from baseline. Baseline values were 1.29 and -0.01 respectively.}
\label{fig:diffConditions}
\end{figure}

Deepfake detection experiments with human participants typically present videos under conditions that are favorable for detection. However, these conditions do not reflect the reality of encountering a deepfake video while browsing the internet. How do deepfake detection rates change under more ecologically-valid conditions? Are there any conditions that detection rates remain robust to?

We tested the detectability of deepfakes across 5 different viewing conditions. The Baseline condition used settings similar to most deepfake detection studies to date: participants performed an unspeeded two-alternative forced choice task, where they viewed one video at a time and reported whether it was real or fake (Figure 1A). Half of the videos were fake (i.e. 50\% prevalence). Data were collected in online experiments, conducted on the Prolific platform (www.prolific.co).

The remaining conditions approximated some of the challenges that might arise in a typical browsing session. This included: 1) Low-Prevalence, where only 20\% of videos were deepfakes, 2) Brief Presentation, where participants were only shown 2 seconds of the video, mimicking a situation where only a brief clip of video is provided or attended to, 3) Divided Attention, where we simulated multitasking by asking participants to perform a digit counting task at the same time as the detection task, and 4) Noisy Video, where the video was blurred and degraded, mimicking situations where video quality is reduced due to compression and streaming limits \cite{prasad2022human}. These conditions targeted both endogenous factors under the control of the user and exogenous factors that may depend on the browsing session.

Stimuli consisted of 360-by-360px videos of faces, selected and cropped from the Deepfake Detection Challenge dataset preview version (\cite{dolhansky2019deepfake}, see Methods and Materials). There were 180 participants per experiment, for a total of 900 participants. Results were analyzed in a signal detection framework \cite{abdi2007signal, batailler2022signal}, with a particular focus on hit rates (i.e. proportion correct on target present trials), since this is the most direct measure of how many deepfakes were correctly detected under different conditions. Additionally, this avoids ambiguity that can arise from only reporting overall accuracy rates: overall accuracy does not distinguish whether participants were good at identifying which videos were fake (i.e. correct on target present trials) or at confirming which videos were real (i.e. correct on target absent trials). As an extreme example, in a setting with 50\% target prevalence, a participant who thought all videos were real would be correct on half of the trials, even if they missed 100\% of the deepfakes. This distinction is especially important in low prevalence settings, when a high rate of correct rejection can make overall accuracy high, even in the face of a low hit rate.

Results for the detection experiments are reported in Figure 1C (and also in tabular format in the Supplement). In the Baseline condition, the average hit rate across participants was 73.3\% (sem (standard error of the mean): 11.3\%), at the cost of a relatively high false alarm rate (28.1\%). Crucially, the hit rate was reduced for all of the experimental conditions we examined. Low Prevalence suffered the most, with the average hit rate reduced to 54.8\% ( sem: 15.8\%), followed by Brief Presentation and Noisy Video (65.2\% and 65.5\%, sem: 10.7 and 11.4 respectively) and Divided Attention (67.4\%, sem: 12.8). All differences were significant, with effect sizes in the medium-large range (Low Prevalence: p=9.44, t$<$0.001, d\textsubscript{s}=1.43; Brief Presentation: p=5.13, t$<$0.001, d\textsubscript{s}=0.77; Divided Attention: p=3.39, t$<$0.001, d\textsubscript{s}= 0.51; Noisy Video: p=4.78, t$<$0.001, d\textsubscript{s}=0.72). Altogether, these results suggest that deepfake detection rates are sensitive to the conditions under which detection is occurring, and that detection is lowered when deepfakes are rare, when engagement with the video is short, when participants are distracted, or when the video is degraded. 

What accounts for these reduction in hit rates? A benefit of the signal detection approach is that it can measure multiple components of a response process. The first component is sensitivity, which measures the subjective perceptual difference between target absent and target present stimuli. The second is criterion, which is the amount of signal that must be present for an observer to make a "target present" response. Interestingly, the mechanism for the reduced hit rate differs across viewing conditions (Figure 1D). For Brief Presentation, Divided Attention and Noisy Video, there is a significant decrease in sensitivity relative to Baseline (Brief Presentation: p=3.21, t=0.0016, d\textsubscript{s}=0.48; Divided Attention: p=3.04, t=0.0027, d\textsubscript{s}= 0.46; Noisy Video: p=9.27, t$<$0.001, d\textsubscript{s}=1.39; Low Prevalence:  p=1.14, not significant) suggesting that the perceptual difference between real and fake videos are not as salient when people are rushed, distracted, or when the video quality is reduced. In contrast, for Low Prevalence, and to a lesser extent for Brief Presentation, there is a substantial increase in criterion (Brief Presentation: p=11.0, t$<$0.001, d\textsubscript{s}=1.66; Low Prevalence: p=3.22, t=0.0015, d\textsubscript{s}= 0.48; Noisy Video: p=1.01, not significant; Divided Attention:  p=1.36, not significant). This suggests that participants are biased to respond “REAL” when they are rushed or when fake videos are rare, and require more obvious signs of tampering in order to overcome this bias. Taken together, different viewing conditions affect different dimensions of detection performance.

So far we have discussed how individuals perform when confronted with real and fake videos, across a range of conditions. However, it is also valuable to understand the accuracy of group-level responses on individual videos. Certain fake news detection systems rely on crowdsourced ratings and explanations, but it is an open question whether consensus responses to deepfake videos are accurate and resilient to viewing conditions. Recent work \cite{groh2022deepfake} found that consensus among human raters was similar to the leading computer vision model. In an exploratory analysis, we took a similar wisdom-of-crowds approach to our data, and examined aggregate responses on individual videos. For each video, we took the majority response (“REAL” or “FAKE”) on a given video as the “consensus response”, and compared it to the ground truth. In the Baseline experiment, the consensus response was correct 84\% of the time, slightly higher than previous results (74\% and 80\% in \cite{groh2022deepfake}), and >10\% higher than individual accuracy (the average overall accuracy in the Baseline condition was 72.9\%). 

Is this aggregate performance level resilient to ecological viewing conditions? In keeping with the exploratory nature of this analysis, we only report effect sizes, using Cohen’s h, for calculating effect sizes of proportions. In general, changes in viewing conditions had little effect on the accuracy of the consensus response, with accuracies of 82\%, 82\% and 80\% for Low Prevalence, Brief Presentation and Divided Attention, respectively (Cohen’s h: 0.05, 0.05 and 0.01 respectively). The only viewing condition to have an effect was Noisy Video, with a consensus performance of 72\% (Cohen’s h: 0.3, considered small to medium). Taken together, this indicates that while individual performance is susceptible to changes in viewing conditions, aggregate measures are more resilient.

\subsection*{Artifact amplification increases deepfake detectability across viewing conditions}

\begin{SCfigure*}[\sidecaptionrelwidth][t]
\centering
\includegraphics[width=14cm]{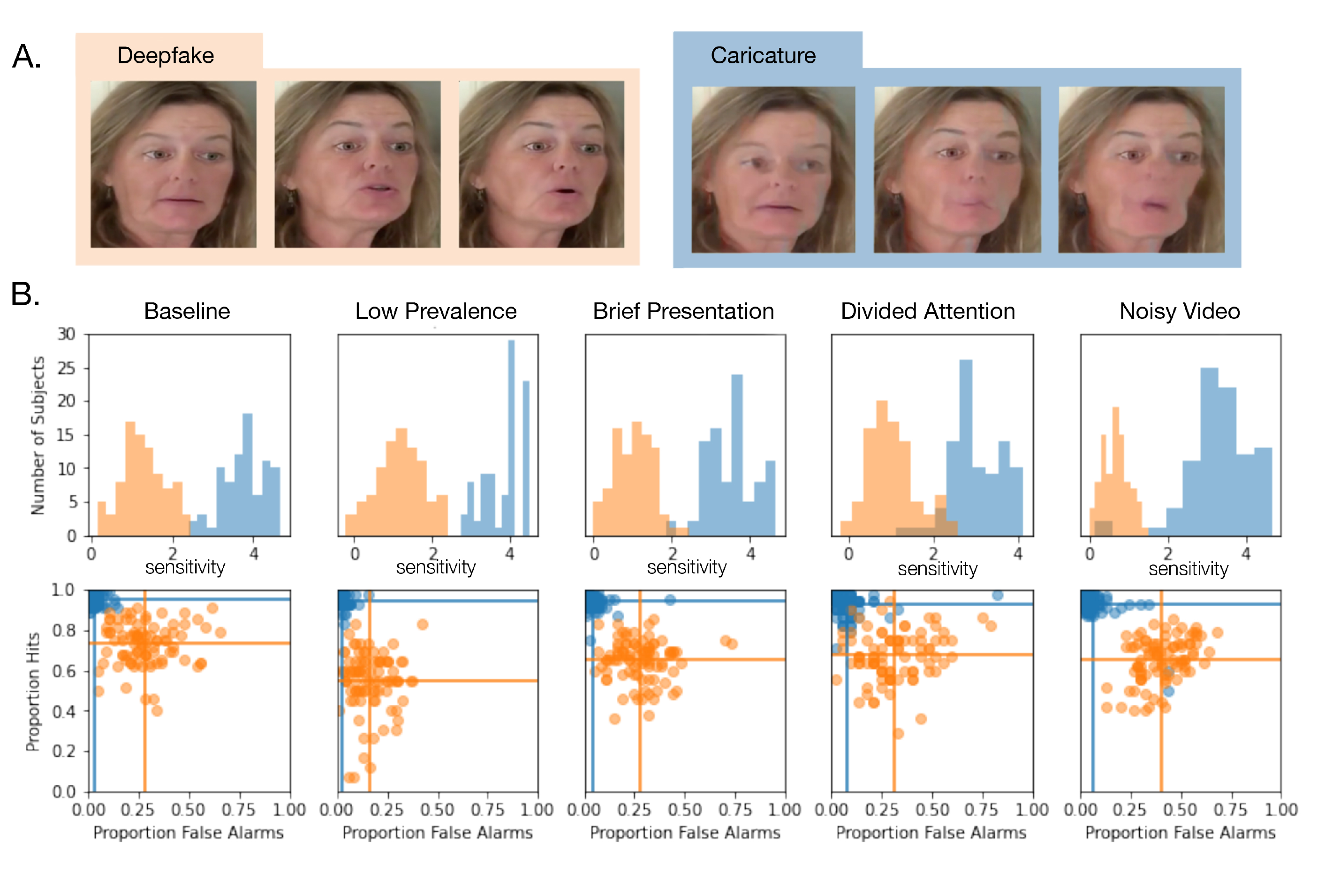}
\caption{Detectability of Deepfake Caricatures. A) Comparison between frames of a deepfake video and the same video with the Caricature transformation applied. The amplification of motion artifacts causes the faces in Caricatures to warp and distort. B) Results: top row shows the distribution of average participant sensitivity under different viewing conditions. Orange denotes participants who saw plain deepfakes, and blue denotes Caricatures. The bottom row replots the data, explicitly showing average participant hit and false alarm rates for deepfakes and Caricatures across viewing conditions.}
\label{fig:cariResults}
\end{SCfigure*}

While human observers can still achieve above-chance success at detecting deepfake videos, this advantage may not last much longer. As deepfakes become more realistic, deepfake mitigation may rely on pairing the human user with a machine learning model. How will these models communicate their predictions to the human user? What kinds of visual indicators will they employ? Here, we test the viability of using artifact amplification for indicating fake videos. Artifacts are strong and intuitive signals that a video is fake, so it is possible that a visual indicator which increases the perception of artifacts may be more detectable and more convincing. 

For these studies, we used a computer vision model that detects then amplifies artifacts in deepfake videos \cite{fosco2022deepfake}. In this approach, the model generates a heat map predicting the locations of artifacts in the input video. In addition to training on large sets of deepfakes, the model is semi-supervised with human annotations of artifacts, so these heat maps identify artifacts that are salient to people as well as machines. This heatmap is used to guide the application of motion magnification to frames of the video, yielding distorted versions of deepfakes where the faces appear to ripple and warp. These distorted outputs are called Deepfake Caricatures (Figure 2A), and we use them to test the effectiveness of artifact amplification as a visual indicator for fake video.

We first established whether making a Caricature of a deepfake improves their detectability. A separate pool of participants (N = 180 per study, for a total of 900) were recruited to perform detection tasks as above, except all fake videos had been subjected to the Caricatures transformation. To quantify how much Caricatures facilitate detection, we report the difference in average sensitivity between participants viewing plain deepfakes (using the above data) and those viewing Caricatures, for a given detection setting. Across all conditions, Caricatures led to a substantial increase in sensitivity (t>20, p<1e\textsuperscript{-50} for all conditions, see Supplement for full statistical reporting). The hit rate was improved to 95.1\% in the Baseline condition, and importantly, remained high across all other conditions, with hit rates of 94.9\%, 94.4\%, 92.6\% and 93.2\% for Low Prevalence, Brief Presentation, Divided Attention and Noisy Video respectively (Figure 2B, see Supplement for all signal detection measures). In all cases, the distribution of sensitivity scores across participants in the Caricature condition had little to no overlap with the distribution for participants in the Deepfake condition, indicative of very large effect sizes (Cohen’s d\textsubscript{s}= 4.67, 4.79, 4.41, 3.41, 4.57 for Baseline, Low Prevalence, Brief Presentation, Divided Attention and Noisy Video, respectively).

Taken together, these results show that artifact amplification is highly effective at making fake video distinguishable from real, and that this increase is present across a range of viewing conditions.

\subsection*{Artifact amplification is more convincing than traditional text-based prompts}

\begin{SCfigure*}[\sidecaptionrelwidth][t]
\centering
\includegraphics[width=14cm]{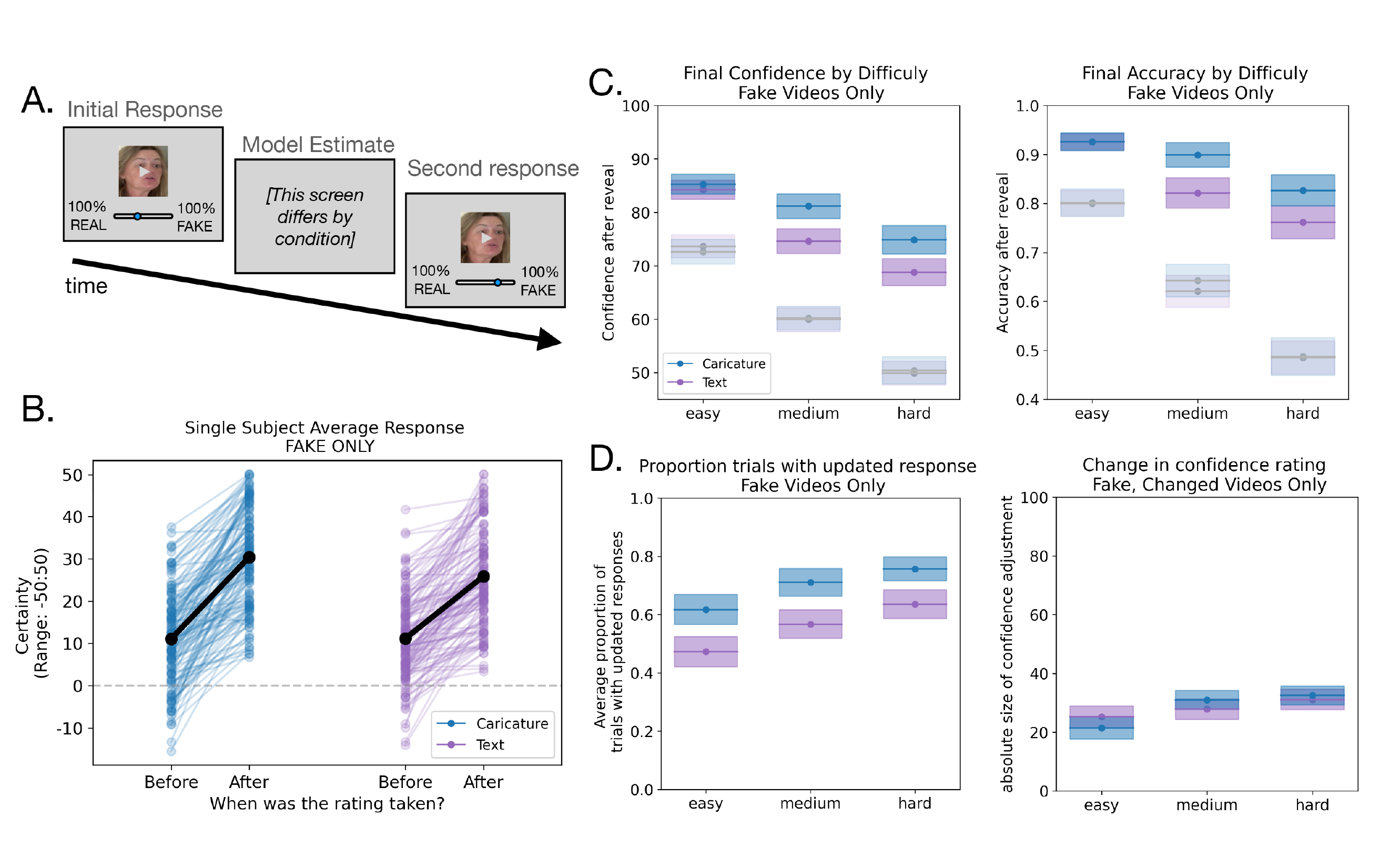}
\caption{ A) Procedure: Participants view a video, make their responses using a slider. Next, they view model prediction. In the Text condition, this screen showed text that read “Our model estimates that this video is REAL” (or “FAKE”). In the Caricature condition, it showed the video with the Caricature procedure applies (this amplifies artifacts in fake videos, but leaves real videos intact). Then, participants could adjust their response. B) Single subject confidence levels, before and after model input, for fake videos. Purple denotes Text, and blue denotes Caricatures. C) Average confidence and accuracy after model input, broken down by deepfake difficulty. Boxes show 95\% CI, and lighter colors show responses before model input. D) Which behavioral changes underlie increased confidence following Caricatures.}
\label{fig:confidence}
\end{SCfigure*}

So far, we have shown that human observers are well below the ceiling at detecting deepfakes, and suggested that artifact amplification is an effective way to boost the detectability of fake videos. However, the perceptibility of a visual indicator is only one way to quantify how effective it is. A crucial second measure of an indicator's effectiveness is whether users find it convincing enough to accept the model’s suggestion.

An ongoing challenge in human-AI teaming is that human users often ignore the suggestion of a model \cite{tschandl2020human, wang2022will, vaccaro2019effects}, even when they report high levels of trust in the model \cite{rechkemmer2022confidence} or are explicitly told that a model is highly accurate \cite{yin2019understanding, boyd2022value}. Deepfake detection models may be particularly susceptible to this, since pictures and videos of people are very compelling, and have been considered a “gold standard” for truth for so long. Indeed, previous work which paired humans and models in deepfake detection tasks found low model acceptance rates \cite{boyd2022value, groh2022deepfake}.

We next examined whether using text-based prompts versus artifact amplification changes the likelihood of model acceptance. We examined three measures of model acceptance : 1) the user’s average degree of confidence in their final, model-assisted response; 2) the proportion of time users accepted the model’s suggestion; and 3) the amount of confidence change participants reported on single trials.

Following Groh et al. (2022), participants were shown a video and provided an initial response on a slider ranging from “100\% confident REAL” to “100\% confident FAKE” (Figure 3A). Next, participants were shown a model prediction screen, where model predictions were conveyed either in text (e.g. “Our model estimates that this video is [REAL/FAKE]”), or by displaying a Caricature of the video. Since the Caricature model works by detecting and amplifying artifacts in fake videos, this has the effect of distorting fake videos and leaving real videos intact. Predictions in this stage were not generated by a real model, but instead reflected ground truth, yielding a “model” performance of 100\% accuracy. This allowed us to collect an estimate of model acceptance in the best-case scenario of a perfect model, and to isolate the role of the visual indicator, since previous work has indicated that model accuracy has its own influence on model acceptance rates \cite{naujoks2016cooperative, sendelbach2013alarm, yin2019understanding}. After viewing the model prediction screen, participants were returned to the screen with the video and given the opportunity to update their response on the slider. 

We hypothesized that the difference between visual indicators might be more pronounced for more difficult videos, which appear more convincingly real. Thus, we included videos at three levels of difficulty (operationalized as their overall detectability in the Baseline condition of the above experiments, see  Methods and Materials). Deepfakes were present at a 50\% prevalence, and there was no time limit for responses. Since this experiment is concerned with how well the different methods convince users that a video is fake, we discuss only target-present trials, where the video is fake.

Figure 3B visualizes a high-level summary of participant’s behavior on target present trials, graphed as the average scores assigned for the video before and after participants viewed model feedback. The score was registered on a 100-point scale centered on 0, where 0 means that participants were unsure if the video was real or fake, 50 means they were sure the video was fake, and -50 means they were sure the video was real. A first observation is that there are large between-subject differences, in both deepfake detection ability (shown by spread of dots in Before condition), and in model acceptance tendencies (shown by the variety in slope between Before and After). Second, it is clear that both visual indicators are effective at changing people’s judgments of the video’s authenticity. 

To quantify the difference in model acceptance between text-based indicators and Caricatures, we examined the difference in the average confidence on After trials across difficulty levels (Figure 3C). Overall, confidence decreased with video difficulty, suggesting that people are more susceptible to challenging deepfakes, even with high-quality model support. Crucially, this decrease is less pronounced in the Caricatures condition: while easy trials showed no difference in final confidence between the two methods, medium and hard trials showed higher subjective confidence for Caricatures (significant interaction: F(2, 798) = 3.51, p = 0.030). Importantly, there was no difference between Text and Caricatures in participant’s responses before model input (grey bars Figure 3C). Thus, models that used artifact amplification to convey their prediction made people more confident in their model-supported decision compared to models that used text-based prompts.

Accuracy scores were also calculated, by transforming the binary confidence score into a binary response. Positive scores were translated to a “Fake” response, negative scores were translated to a “Real” response, and the accuracy of these binary responses were assessed against the ground truth. We found that accuracy scores followed the same pattern as confidence scores (Figure 3C): accuracy of AI-assisted responses were not different between Caricatures or Text for easy trials, but Caricatures led to higher accuracy for medium and difficult trials (significant interaction, F(2,798) = 4.63, p = 0.010). This illustrates how models which elicit higher subjective confidence can increase detection outcomes, in cases where the model is highly accurate.

What change in user behavior underlies this increase in overall confidence? One possibility is that Caricatures increase the frequency with which users accept the model’s suggestion. Another is that model acceptance rates remain the same, but Caricatures lead to larger changes in confidence levels. We compared these options in a follow-up analysis (Figure 3D), and found that the average proportion of trials on which subjects changed their responses was higher for Caricatures, at all levels of difficulty (significant main effect: F(1,798) = 17.28, p<0.001, no interaction effect). In contrast, there was no difference in the magnitude of the adjustment people made following caricatures vs text-based indicators (no main effect: F(1,768) = 2.208813, p = 0.14, no interaction effect; analyzing subset of trials where participants adjusted their responses). This suggests that the efficacy of Caricatures comes from their ability to make an impression more often, not necessarily from their ability to make a larger impression.

\subsection*{Post-hoc individual differences}

\begin{figure}
\centering
\includegraphics[width=1\linewidth]{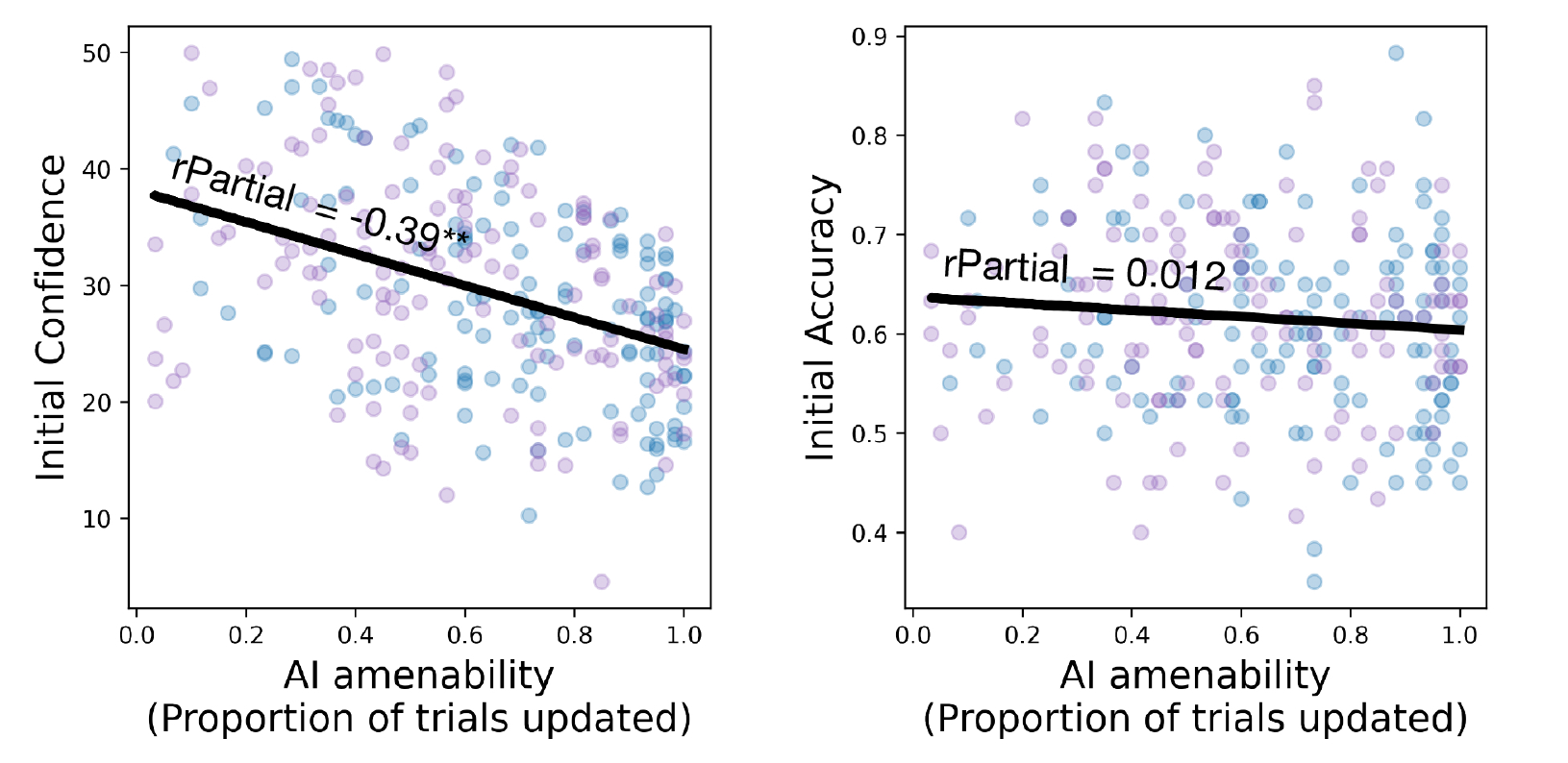}
\caption{Exploratory analysis of behavioral measures that correlate with a participant’s AI amenability, operationalized as the proportion of trials on which they updated their response based on model input. We examined its relationship with participant confidence (before model input) and participant accuracy (before model input). Since confidence and accuracy are related (VIF = 11.96), we used partial correlations. There was no difference in the trend between Text and Caricature trials, so all data were combined for this analysis, but condition colors are preserved in these figures for transparency. Stars indicate significance.}
\label{fig:indivDiff}
\end{figure}

Given the large number of participants, and high inter-individual variation, we saw an opportunity for an exploratory analysis on the individual characteristics that might influence model acceptance rates. We assigned each participant an AI amenability score, operationalized as the proportion of trials on which they adjusted their response following the AI’s feedback, and examined what other behavioral measures were correlated. 

One possibility is that individuals with low AI amenability were simply the individuals where were better at the task: if their initial answers tended to be more correct, there would be no reason to engage with AI assistance. Another possibility is that low-amenability participants were more confident in their initial response, and therefore less likely to adjust, even when the model indicates they were wrong. Of course, a participant’s confidence depends partially on their perceived accuracy in the study. Indeed, the VIF (variance inflation factor) between a participant's initial accuracy (e.i. before model input) and their initial confidence was above 10, indicating high covariance (VIF = 11.96). Thus we used partial correlations, to assess the relationship between AI amenability, and initial accuracy and confidence, respectively, while partialling out the dependence between initial accuracy and initial confidence. 

Overall, results (Figure 4) indicate that there is a a small but significant negative relationship between a participant’s initial confidence (independent of their actual accuracy) and their likelihood to accept the AI’s suggestion (rpartial = -0.39, p<0.001). In contrast, there was no relationship between a participant’s unassisted accuracy and their AI amenability (rpartial = 0.012, p=0.84). This suggests that another possible predictor of whether a user will successfully pair with an AI, independent of the visual design of the indicator, is how confident they feel about their ability to perform the task unassisted. Future work is required to quantify the contribution of this factor.

\section*{Discussion}
Two pressing questions in today’s media landscape are how susceptible are people to deepfakes, and how to mitigate the risks that they pose. Here, we advance our understanding of both of these issues. We show that deepfake detection rates are highly sensitive to the conditions under which they are viewed, and are negatively impacted by many of the conditions present in typical browsing sessions. We also confirm previous findings that pairing a human observer with a machine learning model can increase detection rates, but illustrate how the design of the visual indicator supplied by the model can affect the quality of the collaboration. Specifically, we demonstrate the effectiveness of artifact amplification as a visual indicator, both in terms of its detectability across different conditions, and its impact on participant confidence. We next discuss some insights from these studies, and some limitations.

\subsection*{Deepfake detection under ecological conditions}

One broad implication of these results is that the field of behavioral deepfake detection is  overestimating people’s detection rates. There is limited value in discussing precise detection accuracy values, since detection rates will change as the technology improves, and can depend on the type of deepfake used as stimuli \cite{rossler2019faceforensics++}. Indeed, there were several features of our dataset that should be expected to inflate detection rates relative to other papers: our stimuli were focused on the face regions, they were collected from a single source, and participants received feedback after every trial, making it easier for them to learn what to attend to. Instead, we focus on how detection rates changed when the experimental conditions varied: detection rates were reduced relative to baseline in every condition we tested. These results suggest that misinformation mitigation researchers should increase their estimates of people’s susceptibility to deepfakes, and the false messages they may convey.

Additionally, we found that the specific mechanisms of this reduction varied (e.g. increased criterion vs reduced sensitivity). This suggests that different conditions have independent effects, which may stack when the conditions are combined. There are a number of additional conditions that would be present during a browsing session which were not tested here, but could be expected to further impact performance. Videos are sometimes viewed while scrolling, and this motion may disguise motion artifacts in deepfakes. Videos are often embedded in text, or near other images and banners, which may add clutter to the visual display, further dividing attention. Certain types of videos (e.g. news reports, press conferences) may be less zoomed into faces, leading to smaller regions containing distortions. On a more positive note, many real-world videos also contain features which can help reveal that they were fabricated: videos with sound or that try to imitate the mannerisms and expressions of real and well-known individuals are much harder to fake \cite{groh2022human}. More work is required to understand the full range of conditions, and their individual (and combined) impacts on detection rates.

More broadly, it is useful to name and test the variety of detection conditions that exist in typical browsing sessions, because they have implications for how we deploy warnings about deepfakes. First, they suggest that visual indicators for deepfake signaling should be tested under a variety of conditions, to ensure that they remain useful and robust across all anticipated conditions. Second, they suggest useful directions for literacy campaigns about deepfakes. For example, these results can be used to warn people that blurry videos are particularly good at disguising artifacts, and unusual claims being delivered via blurry videos should be viewed with particular suspicion. Finally, it adds complexity to effort to compare human deepfake detection performance to machine detection \cite{groh2022deepfake, rossler2019faceforensics++}. Models are less likely to be susceptible to prevalence effects of manipulations that divide human attention, but may be more susceptible to other conditions. 

Interestingly, our exploratory results suggest that crowd-consensus deepfake detection is much more robust to ecological conditions than individual detection. This suggests that techniques which rely on aggregate annotations can be successful even if individual judgments are less reliable. One example is the use of human annotation data to supervise deepfake detection models \cite{fosco2022deepfake, boyd2022value, gupta2020eyes}. Another is the use of crowdsourcing as part of a real-time deepfake detection pipeline. Misinformation mitigation in some online communities relies on aggregating reports on content that is already in circulation from users themselves. The present results suggest that this approach could be useful for deepfakes in active circulation on platforms with wide and active user bases. More targeted research is require to confirm these exploratory results. 

\subsection*{Insights about visual indicator design}

A second implication of this work is that human-AI teaming, while effective, has unmet potential, and that the choice of visual indicator can influence model acceptance rates. Our work speaks to previous studies where pairing humans with high-performing deepfake detection models achieved performance well below ceiling \cite{groh2022deepfake, lai2019human, boyd2022value}. These studies used model performance values that were high, but not perfect. This is ecological (no current model performs at 100\% accuracy), but it introduces a confound when trying to assess how visual indicator design affects model acceptance: people are less likely to cooperate with models that have made errors in the past \cite{naujoks2016cooperative, sendelbach2013alarm, yin2019understanding}. Here, we fixed model performance at 100\% which serves two roles. First, it allows us to observe how visual indicator design can affect model acceptance rates without possible interactions with source reliability effects. Second, it allows for observation of model acceptance rates in the best possible scenarios. Overall, even when paired with a perfect model, participants achieved performance well below ceiling (64.4\% for Caricatures, aggregated across difficulty levels). This adds to the growing literature about a human acceptance gap in human-AI teaming for deepfake detection.

These results also make the case that visual indicator design is one factor in reducing this gap. We tested one particular kind of visual indicator, artifact amplification, and found that it is detectable under a variety of viewing conditions, and that it affects participants’ subjective impression of the video more than traditional text-based indicators. This adds to previous results from our group showing that artifact amplification on deepfakes is effective at boosting detection for both high- and low- vigilance individuals, and that it is effective after as little as 500ms of exposure \cite{fosco2022deepfake}. There are several possible reasons that artifact amplification is so effective: it increases the amount of motion in the video, which humans are very perceptually attuned to, and it increases the subjective impression of unnaturalness as the faces change shape over time. Future work is required to untangle these two contributions, potentially by testing the effectiveness of artifact amplification on non-face stimuli. 

Overall, artifact amplification can be considered part of a broader family of distortion-based visual indicators. Such visual indicators rely on the conscious, deliberate distortion of an image to enable visual observation of an otherwise invisible signal \cite{le2019seeing, smieja2021motion}. These have existed for some time across a number industrial and civil settings, as a visual aid in quality control applications. For example, motion amplification has been found useful for monitoring vibrations in iron pipes \cite{kupwade2020corrosion} and pedestrian bridges \cite{shang2018multi}, and for visualizing the deformation in wind turbine blades \cite{sarrafi2018vibration} and antique structures \cite{fioriti2018motion}. Motion and color amplification has even been proposed for facilitating the observation of subtle physiological signals like heart rate in infants \cite{wu2012eulerian, balakrishnan2013detecting}. 

Deepfake mitigation measures have only recently begun to recognize the perceptual power of distortion. Some approaches actively inject human-invisible artifacts into real images or video, which cause any subsequent video manipulation to contain large and visible artifacts \cite{chen2021magdr, wang2022anti}. We introduce a complimentary, reactive approach, which identifies and amplifies distortions caused by the deepfake-generation pipeline itself. Distortion-based indicators could also be applied to deepfakes identified via metadata-based detection methods \cite{qureshi2021detecting, neekhara2022facesigns, chan2020combating, alattar2020system, yu2021artificial}, by injecting artifacts into the video stream. Crucially, we have empirically demonstrated the effectiveness of distortion-based visual indicators in deepfake mitigation, 
and this principle can be applied regardless of the method used to identify the deepfake.

\subsection*{Limitations and risks}
There are some limitations to the present work. The present studies use videos that have no sound, and that are predominantly  focused on faces. This matches many kinds of viewing contexts, such as GIFs (Graphics Interchange Format), which do not contain audio, or platforms that have volume off by default. However, there are also contexts that this does not generalize to, such as the experience of watching a full interview with an individual. This context will have many additional streams of information, including the quality of the audio, and the semantics that are communicated via the audio. It is not clear how these additional information streams will interact with visual indicators which rely on distortions applied only on the video stream. However, such contexts open up novel research directions, such as testing distortion-based indicators in the audio domain. 

Additionally, we assessed the effectiveness of the visual indicator based on the accuracy and confidence achieved by the participants. However, these measures present an incomplete picture of how people will interact with the fake media. Future work on visual indicator design should also consider how a given kind of indicator affects downstream memory for the videos, as well as memory for the information they contain. 

This work also raises questions about the risks of distortion-based visual indicators for deep fake signaling. While we show they are effective for signaling deepfakes in the moment, they may harm longer term information literacy goals. If people only see distorted deepfakes, they may not learn what artifacts exist in unsignaled deepfakes. Widespread distribution of distortion-based visual indicators may also cause a criterion shift, where people become less sensitive to the subtle artifacts in regular deepfakes because they have become accustomed to more obvious visual distortions. 

This work also raises general questions about the risks involved in creating more convincing visual indicators. Manipulations to increase engagement with AI can end up producing over-trust in the AI, which is particularly problematic if the model has low accuracy. In many human-AI teaming situations, there is a “rebound effect”, where users start ignoring the model when they notice errors \cite{cvach2012monitor, ancker2017effects, hussain2019medication}. Is the rebound effect bigger when the visual indicator is more compelling? Finally, are more compelling visual indicators at higher risk of being weaponized to erode trust in real videos? Altogether,  it is an open question whether indicators that have stronger positive effects also have stronger negative effects.

\section{Conclusion}
To conclude, we demonstrate how conditions that exist during normal browsing can increase human susceptibility to deepfakes. However, we also demonstrate how human-centered principles can be applied to visual indicator design to increase their effectiveness. We leveraged people's natural sensitivity to distortions in faces by amplifying artifacts in videos, and found that this method of marking fake videos was more convincing than text-based alerts. More broadly, this paper demonstrates the promise of integrating knowledge about what perceptual tasks are easy and automatic for humans into the development of visual indicators.

\matmethods{

\textbf{Pre-registration.} All analyses were pre-planned, except where they are described as exploratory or "follow-up" in the text. All experiments were internally pre-registered, and the following experiments had pre-registration posted on AsPredicted.org:
Baseline, Low Prevalence, Brief Presentation, Divided Attention, 
Noisy Video (https://aspredicted.org/63L\textunderscore W8X, https://aspredicted.org/VX6\textunderscore TDZ, https://aspredicted.org/132\textunderscore 1CL, https://aspredicted.org/ZV1\textunderscore 2JY, https://aspredicted.org/VKK\textunderscore W35)

\textbf{Study 1: Detectability of Deepfake Caricatures}

\textit{Stimuli.} Stimuli consisted of videos of single individuals. Videos were selected from the Deepfake Detection Challenge preview dataset (DFDCp  \cite{dolhansky2019deepfake}. Videos were pre-processed as described in \cite{fosco2022deepfake}. Sounds were removed, videos were cut into 12 seconds clips and cropped to only show one face. Each cropped clip measured 360x360 pixels, with a minimum of a 100px margin between the edge of frame and the face. We created a set of 900 video clips, which included 300 real videos, the corresponding 300 deepfakes, and the corresponding 300 Caricatures generated from these deepfakes.

First, we selected 300 real clips, by sampling 2-5 real video clips for different actors in the DFDCp. We sampled the videos to include variation in gender, race, body type, hair, age, and bearing. Next, we selected the corresponding deepfakes. The DFDCp features multiple deepfakes generated from each real video. For each of the real clips we selected, we retrieved all of the corresponding deepfakes, and filtered them for quality. Our goal was to match the quality of deepfakes in our study to the quality of deepfakes that would plausibly be shared online. Thus, we excluded any deepfake which contained artifacts for its whole duration, contained artifacts covering the whole face at any point in time, contained a momentary failure revealing the real face underneath, had mismatches in gender, lighting or skin tone from the underlying head and body. We additionally removed any deepfake that was indistinguishable from the real face it was generated from. For each real clip, we randomly selected one of the corresponding deepfake clips from the set that survived this filtering, yielding 300 real-fake pairs.

Finally, from each of the tampered videos, a Caricatures was created using the CariNet approach\cite{fosco2022deepfake}. CariNet is a semi-supervised framework that predicts which regions in a tampered video of a face contain artifacts that are salient to human observers, and selectively amplifies them using motion magnification. 

Our experiments were calibrated to take a median of 15 minutes (in the Baseline condition, see below). Thus, our set of 300 videos was subdivided into folds, and presented to subjects such that each subject saw 50 real videos and 50 fake videos (fake videos within a fold could not be generated from real videos in the same fold). 

\textit{Sample size estimation.} Sample size was determined by estimating the number of participants required to gain a stable estimate of the detectability of a single video, based on pilot data. A pilot study was conducted with 40 participants per condition, using the Baseline procedure (see below). Following \cite{strong2019using}, we simulated sample sizes ranging between 10 and 200, by sampling 1000 times with replacement from the pilot participants. For each sample size, we calculated the variance in average detectability for a given video, averaged over all videos. We identified the sample size at which this variance plateaus (N = 6), then doubled and rounded up to obtain a sample size of 15 subjects per video. 

\textit{Participants}. A total of 913 people participated across all sub-experiments in Experiment 1 (52\% Female, 3\% percent Not Reported). Participants were recruited from the Prolific online experiment platform. Participants were required to meet the following criteria: 95\% approval rates or higher, a history of more than 500 tasks completed on the platform, and located in the US. Participants were recruited and compensated according to procedure approved by MIT’s Committee on the Use of Humans as Experimental Subjects. Participants were paid an hourly rate of \$11.25 per hour.

\textit{Design.} There were five slightly different designs of detectability experiments, depending on the viewing condition being measured. 

In the Baseline condition, participants viewed one video at a time and indicated whether they thought the video was real or fake. Responses were not time limited, and participants received feedback on every trial. There were 100 trials, divided into five blocks of 20. The experiment started with five exposure trials, in which a 12-second deepfake video was displayed, but no response was required. The experiment contained five attention check trials, which consisted of a video with the message “ this is an attention check, please select “REAL” ”  in capital letters. Real and fake videos were equally prevalent, and randomly intermixed.  

In the Low Prevalence condition, the same design was used, with the exception that only 20\% of the trials contained fake videos. The subset of deepfakes used here were randomly sampled from the full set. 

In the Speeded Presentation condition, the Baseline condition was modified such that videos were only presented for 2 seconds. This value was selected by taking the median reaction time in the Baseline condition (2.8s), then rounding down to yield a moderately challenging time limit. In order to enforce the time limit, the response screen appeared after the video disappeared.

In the Divided Attention condition, participants performed a digit counting task. Similar tasks have been used to reduce participant bandwidth in driving research \cite{yamani2018impact, horrey2006modeling}. A nine-digit string was displayed on the video itself, one digit at a time, with a 0.45 second interval between digits. Participants were asked to count the number of odd digits in the string, which ranged from 3 to 5 (inclusive). Participants reported the number of digits after reporting their response for the video. Because this is a challenging task, the experiment was shortened to 50 trials. 

The Noisy Video condition was identical to the Baseline condition, with the exception that the videos had been manipulated to mimic compression artifacts due to lossy encoding. Similar to \cite{rossler2018faceforensics}, videos were compressed using a constant rate factor of 40 (18 is considered perceptually lossless, 23-28 is considered acceptable), yielding blurring and aliasing.

Each of these viewing conditions had a deepfake version, or a caricature version. These versions were presented in a between-subjects design, because we were concerned that including deepfakes and caricatures in the same subject would cause criterion shifts. We took the following steps to reduce population effects: there were no outward differences between the deepfake and caricature versions until participants started the task, both versions of the task were released on the website at the same time, and condition assignment was simply determined by which link participants clicked. 

\textit{Analysis.}
The following pre-registered procedures were used for removing low quality data: any participant who failed three or more vigilances were removed and replaced, and any trial that took longer than 60 seconds was dropped. For the divided attention condition, we additionally dropped any subject performing lower than two standard deviations below the mean on the digit counting task, in order to ensure they were devoting sufficient attention to the number task.

Some participants had 100\% accuracy rates, especially in the caricatures condition. Thus,  for calculating calculating signal detection measures, we used the method proposed in \cite{hautus1995corrections} to calculate sensitivity and criterion in case with extreme values (.5 is added to the count of Hits, False Alarms, Misses and Correct Rejections to avoid a divide by zero).

Power analysis indicated that we had very high power, so we took a conservative approach to quantifying differences  in hit rate, false alarm rate, sensitivity and criterion from Baseline.  Specifically, we used a Bonferroni correction where the alpha level was divided by the total number of tests (16). Effect sizes were measured using Cohen’s ds, for two-sample designs \cite{lakens2013calculating}. To compute the statistical significance of the sensitivity difference within each condition, for which we pre-registered a hypothesis, we used one-sided unpaired t-test, with a Bonferroni correction such that alpha level was divided by the total number of tests (one for each experimental condition, 5). 

\textbf{Study 2: Caricatures compared to text-based prompts}

\textit{Stimuli.} Stimuli for Study 2 were a subset of those in Study 1. Deepfakes from Study 1 were given a detectability score, which was the percent of time it was detected in the Baseline condition, across all participants who viewed it. Overly easy and difficult deepfakes were discarded (detection rates below 20\% or above 90\%), and the remaining deepfakes were binned by difficulty: difficult (detected 0.21\% to 0.44\% of the time), medium (0.44\%-0.67\%), and easy (0.67\%, 0.9\%). Finally, 30 deepfakes were randomly selected from each bin to make up the stimulus set, as well as the corresponding caricatures. To achieve 50\% fake prevalence, and additional 90 real videos were randomly selected, with the caveat that they not include the real videos corresponding to the selected deepfakes.

\textit{Participants.} Participant recruitment followed the same criteria, ethical safeguards, and payment rates as Experiment 1. A total of 298 people participated (47\% Female, 1\% percent Not Reported). 

\textit{Design.} Our design was based on the design in Groh (2021) \cite{groh2022deepfake}. Participants saw one video at a time, and were asked to “Rate your certainty that this is a deepfake” using a slider, whose poles were labeled “100\% confident REAL” and “100\% confident FAKE”. Next, they clicked a button labeled “Click here to see the model's guess”, and were shown the prediction of the model as either a text-based prompt or a caricature. 

Text-based prompts were of the form “the model estimates that this video is REAL” (or “FAKE”). They were introduced with the text “Here is the output of the modeling procedure. As a reminder, our model works by estimating the likelihood that the video is fake, then labeling the videos it thinks are probably fake”.
In the Caricatures, this screen displayed the same video after it had been passed through the Caricature model, which  distorts deepfakes but leaves real videos intact. The caricatures were introduced with the text “Here is the output of the modeling procedure. As a reminder, our model works by estimating the likelihood that the video is fake, then distorting the videos it thinks are probably fake”. 
Real-fake labels reflected the ground truth.

This experiment included real and fake at 50\% prevalence, there was no time pressure, and participants received feedback on their response on each trial. There were 60 trials total per participant, divided into blocks of 10. There were 5 randomly placed catch-trials, on which the text “this is an attention check, please set confidence to 100\% Real” was displayed instead of the model prediction. Text-based and Caricature conditions were collected in a between-subjects manner. 

\textit{Data analysis}
The following pre-registered procedures were used for removing low quality data: any participant who failed three or more vigilances were removed and replaced, and any trial that took longer than 60 seconds was dropped.

Four analyses were performed on this data, each using an ANOVA to test for a main effect of visual indicator type (text-based prompt or caricature) or interaction between visual indicator type and difficulty on each of four measures of interest (if an interaction was present, main effects were not analyzed). There were two pre-planned measures of interest (final confidence and final accuracy), for which we used standard p value of 0.05, and two post-hoc measures (proportion of trials on which participants updated their responses, magnitude of response changes) for which we used a Bonferoni-corrected p-value of 0.025. 

\textbf{Post hoc individual differences} Single subject averages were extracted for initial accuracy (i.e. accuracy before model input), initial confidence (e.i. confidence level before model input) and AI amenabilty (i.e. the proportion of trials a participant updated their responses following model feedback). Partial correlations for the individual difference measures were performed using the ppcor package in R.}

\showmatmethods{}

\acknow{This work was partially funded by an Ignite grant from SystemsThatLearn@CSAIL, and made possible by generous funding from Eric Yuan, CEO and founder of Zoom Video Communications}
\showacknow{} 

\bibliography{pnas-bib}

\end{document}


\maketitle
\thispagestyle{firststyle}
\ifthenelse{\boolean{shortarticle}}{\ifthenelse{\boolean{singlecolumn}}}

\section{Signal detection measures for non-signaled deepfakes across viewing conditions}
\begin{tabular}{||c c c c c||} 
 \hline
 \textbf{Viewing Condition} & \textbf{Hit Rate} & \textbf{False Alarm Rate} & \textbf{Sensitivity} & \textbf{Criterion} \\[0.5ex] 
 \hline
 Baseline           & 0.73 & 0.28 & 1.29 & -0.01 \\ 
 \hline
 Low Prevalence     & 0.55 & 0.16 & 1.19 & 0.48 \\
 \hline
 Brief Presentation & 0.65 & 0.28 & 1.04 & 0.11 \\
 \hline
 Divided Attention  & 0.67 & 0.31 & 1.03 & 0.03 \\
  \hline
 Noisy Video        & 0.66 & 0.40 & 4.57 & -0.08 \\ [1ex] 
 \hline
\end{tabular}

\section{Complete statistical reporting of sensitivity difference between non-signaled deepfakes, and deepfakes signaled using Caricatures}
\begin{tabular}{||c c c c||} 
 \hline
 \textbf{Viewing Condition} & \textbf{t} & \textbf{p} & \textbf{Effect Size} \\[0.5ex] 
 \hline
 Baseline & 30.08 & 2.50E-69 & 4.66 \\ 
 \hline
 Low Prevalence & 31.86 & 3.14E-75 & 4.79 \\
 \hline
 Brief Presentation & 29.62 & 1.00E-70 & 4.41 \\
 \hline
 Divided Attention & 22.73 & 1.56E-54 & 3.41 \\
  \hline
 Noisy Video & 22.73 & 9.84E-76 & 4.57 \\ [1ex] 
 \hline
\end{tabular}

\section{Signal detection measures for deepfakes signaled with Caricatures across viewing conditions}

\begin{tabular}{||c c c c c||} 
 \hline
 \textbf{Viewing Condition} & \textbf{Hit Rate} & \textbf{False Alarm Rate} & \textbf{Sensitivity} & \textbf{Criterion} \\[0.5ex] 
 \hline
 Baseline           & 0.95 & 0.03 & 3.75 & 0.13 \\ 
 \hline
 Low Prevalence     & 0.95 & 0.02 & 3.85 & 0.22 \\
 \hline
 Brief Presentation & 0.94 & 0.04 & 3.07 & 0.10 \\
 \hline
 Divided Attention  & 0.93 & 0.08 & 3.06 & -0.009 \\
  \hline
 Noisy Video        & 0.93 & 0.06 & 3.32 & 0.05 \\ [1ex] 
 \hline
\end{tabular}

\vspace{1cm}

The figure below shows a graphical representation of differences from baseline across viewing conditions for Caricatures. Light blue box indicates standard error of the mean, and stars indicate significance (post-hoc Bonferroni corrected two-sample t-test). In contrast to non-signaled deepfakes, viewing conditions only reduce hit rates in one case, when the viewer is distracted.

\begin{SCfigure}
\includegraphics[width=0.75\linewidth]{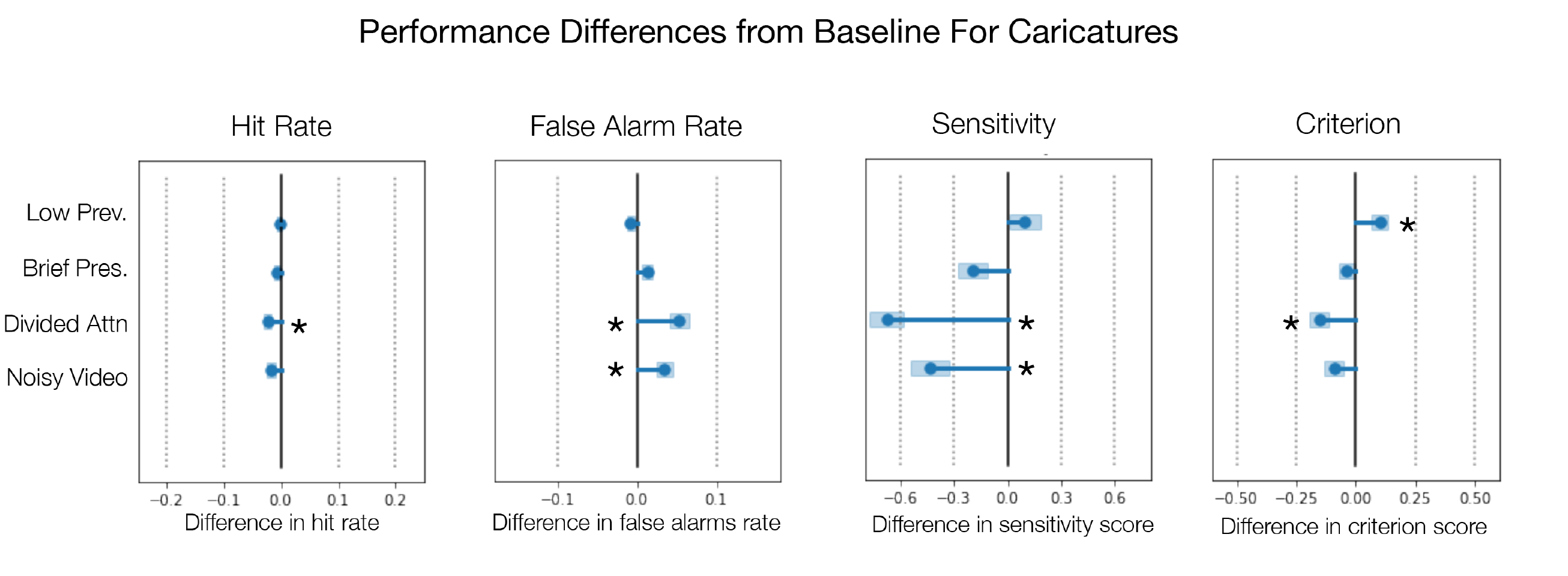}
\label{fig:Supp1}
\end{SCfigure}

\bibliography{pnas-bib}